\documentstyle[epsfig,12pt]{article}

\newcommand{\bce}{\begin{center}} 
\newcommand{\ece}{\end{center}}
\newcommand{\beq}{\begin{equation}}
\newcommand{\eeq}{\end{equation}}
\newcommand{\bea}{\vspace{0.25cm}\begin{eqnarray}}
\newcommand{\eea}{\end{eqnarray}}

\newcommand{\ba}{\begin{array}}
\newcommand{\ea}{\end{array}}

\newcommand{\doublespace}{
    \renewcommand{\baselinestretch}{1.6}\large\normalsize}

\def\lsim{\mathrel{\rlap{\lower4pt\hbox{\hskip1pt$\sim$}}
    \raise1pt\hbox{$<$}}}         
\def\gsim{\mathrel{\rlap{\lower4pt\hbox{\hskip1pt$\sim$}}
    \raise1pt\hbox{$>$}}}         

\def\Pom{{\bf I\!P}}

\def\beq{\begin{equation}}
\def\endeq{\end{equation}}
\def\arr{\begin{eqnarray}}
\def\endarr{\end{eqnarray}}
\makeindex

  \advance\oddsidemargin  by -1in
  \advance\evensidemargin by -1in
  \advance\topmargin      by -0.5in
\begin{document}

\vspace{2.0cm}

\begin{flushright}
\end{flushright}

\vspace{1.0cm}

\begin{center}
{\Large \bf Beauty, charm and $F_L$ at HERA:\\ 
new data vs. early predictions.
\vspace{1.0cm}}

{\large\bf N.N. Nikolaev$^{1 \dagger}$ and 
V.R.~Zoller$^{2 \ddagger}$}

\vspace{1.0cm}

$^{1)}${ \em
Institut  f\"ur Kernphysik, Forschungszentrum J\"ulich,
D-52425 J\"ulich, Germany\\
and\\
L.D.Landau Institute for Theoretical Physics, Chernogolovka,
142432 Moscow Region, Russia}\\

$^2${\em ITEP, Moscow 117218, Russia}

\vspace{1.0cm}

{\bf Abstract}
\end{center}
One of the well known effects of the asymptotic freedom is  splitting of
  the leading-$\log$ BFKL pomeron
 into a  series of isolated poles in complex angular  momentum plane.
 Following our earlier works  we explore the phenomenological 
consequences 
of the emerging BFKL-Regge factorized expansion for the small-$x$  
charm ($F_2^c$) and beauty ($F_{2}^{b}$) structure functions  of the proton.
As we found earlier,the color dipole approach to the BFKL dynamics
 predicts uniquely decoupling 
of subleading hard BFKL exchanges from $F_2^c$ at moderately large 
$Q^{2}$. We predicted precocious BFKL asymptotics of $F_2^c(x,Q^2)$
 with intercept of the rightmost  BFKL pole 
$\alpha_{\Pom}(0)-1=\Delta_{\Pom}\approx 0.4$. 
High-energy open beauty
photo- and electro-production probes the vacuum exchange at much smaller 
distances  and detects significant  corrections to the 
BFKL asymptotics coming from the  subleading  vacuum poles.
In view of the  accumulation of the experimental data on small-$x$
$F_{2}^{c}$ and $F_{2}^{b}$  we extended 
our early predictions to 
the kinematical domain covered by new HERA measurements.
Our structure functions obtained  in 1999
 agree well  with the  determination of both $F_2^c$ and $F_2^b$  
by the H1 published in 2006 but  contradict to very recent (2008, preliminary) 
H1 results on $F_2^b$.  We present also  comparison of our early
 predictions  for  the longitudinal structure function $F_L$ 
 with recent  H1 data (2008) taken at very low Bjorken $x$. We 
comment on the electromagnetic corrections to the 
Okun-Pomeranchuk theorem.

\doublespace

\vskip 0.5cm \vfill $\begin{array}{ll}
^{\dagger}\mbox{{\it email address:}} & \mbox{n.nikolaev@fz-juelich.de} \\
^{\ddagger}\mbox{{\it email address:}} & \mbox{zoller@itep.ru} \\
\end{array}$

\pagebreak


\section{Introduction}

The Okun-Pomeranchuk theorem on the isospin independence of asymptotic cross sections
\cite{OP} is a precursor to the QCD pomeron which is a flavour-neutral bound state of
gluons  in the $t$-channel.  Within the color-dipole (CD)  framework, this flavor
independence is a feature of the dipole cross section, while the QCD pomeron contribution
would depend on the interacting particles through the QCD impact factors, calculable in
terms of the flavour-dependent color dipole structure of the target and projectile.

As noticed by Fadin, 
Kuraev and Lipatov in 1975 (\cite{FKL}, see also more detailed discussion by
Lipatov \cite{Lipatov}), incorporation of asymptotic freedom into BFKL
equation  turns the spectrum of the QCD vacuum exchange into series of
isolated BFKL-Regge poles. 
Such a spectrum has a far-reaching theoretical
and experimental consequences because
the contribution of each isolated hard BFKL pole to scattering amplitudes
and/or  structure functions (SF) would satisfy very powerful 
Regge factorization \cite{Gribov}.
The resulting CD BFKL-Regge factorized expansion allows one to relate in a
parameter-free fashion SFs of different targets, 
$p,\pi,\gamma,\gamma^{*}$ \cite{DER,PION,GamGam} and/or 
contributions of different flavors to 
the proton SF \cite{CHARM2000, BEAUTY}. The first analysis of
small-$x$ behavior of
open charm SF of the proton, $F_2^c$,
 in the color
dipole formulation of the BFKL equation \cite{BFKL}
 has been carried out in 1994 
\cite{NZZ94,NZDelta,NZHERA} with an intriguing result that for moderately
 large 
$Q^{2}$ it is dominated by the leading hard BFKL pole  exchange.
Later this fundamental 
feature of CD
BFKL approach has been related \cite{NZcharm} to nodal properties 
of eigen-functions of
subleading hard BFKL-Regge poles \cite{NZZ97}.

In \cite{NZZ97} we applied the latter
property of the CD BFKL-Regge factorization and quantified the strength of
the subleading hard BFKL and soft-pomeron background to dominant rightmost
hard BFKL exchange. One of our findings \cite{NZZ97} is that the BFKL-Regge
 expansion (\ref{eq:3.6}) truncated at $m=2$ appears to be very successful
 in describing of the proton SFs in a wide range of $Q^2$. Very recently this
phenomenon  has been rediscovered in \cite{ELLIS}.

In view of the  accumulation of the experimental data on small-$x$
$F_{2}^{c}$, $F_{2}^{b}$   we extended 
our early predictions to 
the kinematical domain covered by new HERA measurements.
 Based on CD
BFKL-Regge factorization we report parameter-free description of  both
$F_2^c$ and $F_2^b$. 
  We comment  on the phenomenon of decoupling of soft and
subleading BFKL singularities at the scale of the open charm production
 which results in precocious color dipole
 BFKL asymptotics of the
the structure function $F_2^c$.In view of this
fundamental conclusion open charm excitation by real photons and in DIS 
gives a particularly clean access to the intercept of the rightmost hard 
BFKL pole for which our 1994 prediction has been $\Delta_{\Pom}=
\alpha_{\Pom}(0)-1= 0.4$ \cite{NZZ94}. 

We show that the
interplay of leading and subleading vacuum exchanges gives rise to the
beauty structure function $F_2^b$ growing much faster than it is
prescribed by the exchange of the leading pomeron trajectory with
intercept $\Delta_{\Pom}=0.4$ (see also \cite{BEAUTY}).

 Because the CD BFKL-Regge
expansion for color dipole-proton cross section has already been fixed 
from the related and highly successful phenomenology of light flavor 
contribution to the proton SF the CD BFKL-Regge factorization 
predictions for the charm SF of the proton are parameter 
free. The found nice agreement with the experimental data from H1
Collaboration \cite{H1ccbb} on the charm and beauty
SF of the proton  strongly corroborates our 
1994 prediction 
$\Delta_{\Pom}\approx 0.4$ for the intercept of the 
rightmost hard BFKL pole. It is worth mentioning that very recent (preliminary)  H1 date on $F_2^b$
\cite{H12008} does not agree with our early predictions.

Besides charm and beauty structure functions there are several  more observables 
which are selective
to the dipole size. One of them is 
 the longitudinal structure function of the proton $F_L$. 
 We present the BFKL-Regge factorization results for $F_L$.
 The recent H1 measurements of $F_L$ \cite{H1FL2008} do not contradict to our 
predictions made in \cite{CHARM2000} but they are too uncertain for 
any firm conclusions.


\section{Open charm production: scanning the dipole cross section}
In color dipole (CD) approach to small-$x$ DIS excitation of heavy flavor 
is described in terms of interaction of $q\bar{q}$ color
dipoles in the photon of  a predominantly small size,
\beq
{ 4 \over Q^{2}+4m_{q}^{2}}\lsim r^{2} \lsim { 1\over m_{q}^{2}}\, .
\label{eq:1.1}
\eeq
Therefore, the  heavy flavor excitation at large values of the Regge parameter,
\beq
{1\over x}={W^2+Q^2\over4 m^2_c + Q^2}\gg 1\,,
\label{eq:1.2}
\eeq
is an arguably sensitive probe of 
short distance properties of vacuum exchange in QCD. 

Interaction of color dipole ${\bf r}$ in the  photon with 
the target proton  is described by the beam, 
target and flavor  independent color dipole cross section 
$\sigma(x,{\bf r})$. The contribution of excitation of open 
charm/beauty to photo-absorption cross section is given by color dipole
factorization formula
\beq
\sigma^{c}(x,Q^{2})=
\int dz d^{2}{\bf{r}} 
|\Psi_{\gamma^*}^{~c\bar{c}}(z,{\bf{r}})|^{2} 
\sigma(x,{\bf{r}})\,.
\label{eq:2.1}
\eeq
Here $|\Psi_{\gamma^*}^{~c\bar{c}}(z,{\bf{r}})|^{2}$  is
a probability to find in the photon the $c\bar{c}$ color dipole with the
charmed quark carrying fraction $z$ of the photon's light-cone momentum
 \cite{NZ91}.
 Hereafter we focus on the charm structure
function  
\bea
F_2^{c}(x_{Bj},Q^2)=
{Q^2\over {4\pi^2\alpha_{em}}}\sigma^{c}(x,Q^{2})
\nonumber \\
=\int{ d r^{2}\over r^{2}}{\sigma(x,r) \over r^{2}} W_{2}(Q^{2},m_{c}^{2},r^2)
\,.
\label{eq:2.4}
\eea
 A
detailed  analysis of  the weight function $W_{2}(Q^{2},m_{c}^{2},r^2)$ 
found upon 
the $z$ integration has been carried out in \cite{NZDelta,NZHERA}, we only cite
the principal results: (i) at moderate $Q^{2} \lsim 4m_{c}^{2}$ the weight
function has a peak at $r \sim {1/ m_{c}}$, (ii) at very high $Q^{2}$ the 
peak 
develops a plateau for dipole sizes in the interval (\ref{eq:1.1}).
One can say that for moderately large $Q^{2}$ excitation of open charm 
probes (scans) the dipole cross section at a special dipole size $r_{S}$
(the scanning radius) 
\beq
r_{S} \sim {1/ m_{c}}\,.
\label{eq:2.6}
\eeq
The difference from light flavors is that in contrast to the peak for heavy
charm the $W_{2}$ for light flavors always has a broad plateau which extends 
up to large dipoles $r \sim {1/m_{q}}$.
\begin{figure}[h]
\psfig{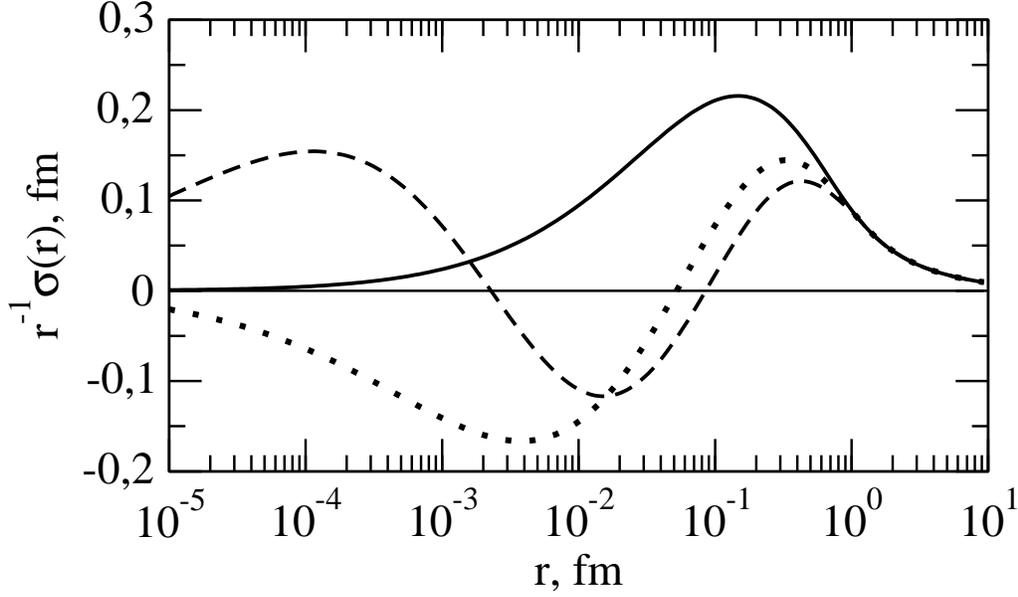}
\vspace{-0.5cm}
\caption{The rightmost $\sigma_0/r$ and subleading $\sigma_1/r$ and  
$\sigma_2/r$
eigen cross sections    as a function of
$r$.} 
\label{fig:fig1}
\end{figure}  

\section{Scanning radius and nodes of subleading CD BFKL eigen-cross sections} 

In the Regge region of ${1\over x} \gg 1$  CD cross section $\sigma(x,r)$
satisfies the CD BFKL equation
\beq
{\partial \sigma(x,r) \over \partial \log{(1/x)}}={\cal K}\otimes \sigma(x,r)\,,
\label{eq:3.1}
\eeq
for the kernel ${\cal K}$ of CD approach see \cite{PISMA1}.
The solutions with Regge behavior
\beq
\sigma_{m}(x,r)=
\sigma_{m}(r)\left({1\over x}\right)^{\Delta_{m}}
\label{eq:3.0} 
\eeq
satisfy the eigen-value 
problem
\beq
{\cal K}\otimes \sigma_{m}=\Delta_{m}\sigma_{m}(r)\,
\label{eq:3.2}
\eeq
and the CD BFKL-Regge expansion for the 
color dipole cross section
reads \cite{NZZ94,PION}
\beq
\sigma(x, r)=\sum_{m=0} 
\sigma_m(r)\left({x_0\over x}\right)^{\Delta_m}.
\label{eq:3.3}
\eeq

 The practical calculation of $\sigma(x,r)$
    requires the boundary condition $\sigma(x_0,r)$ 
at certain $x_0\ll 1$.
We take for boundary condition at $x=x_0$ the Born approximation,
$$\sigma(x_0,r)=\sigma_{Born}(r)\,,$$ 
 i.e. evaluate dipole-proton scattering via the two-gluon exchange.
This leaves the starting point $x_0$ the sole parameter.
We follow the choice $x_0=0.03$ which  met  with  remarkable
 phenomenological success 
\cite{NZZ97,DER,PION}.

The properties of our CD BFKL equation and the choice of
physics motivated boundary condition were discussed in detail elsewhere 
\cite{NZDelta,NZHERA,NZcharm,NZZ97,DER}, 
here we only recapitulate features 
relevant to the considered problem. Incorporation of asymptotic freedom 
exacerbates well known infrared sensitivity of
the BFKL equation and infrared regularization by infrared freezing of the 
running coupling $\alpha_S(r)$ and modeling of confinement of gluons 
by the finite propagation radius of perturbative gluons $R_c$ need to be
 invoked.

The leading eigen-function 
$$\sigma_0(r)\equiv\sigma_{\Pom}(r)$$
for ground state i.e., for the rightmost hard BFKL pole is node free.  
The  subleading  eigen-function for excited state $\sigma_m(r)$ has $m$ nodes. 
We find $\sigma_m(r)$ numerically \cite{NZZ97,DER}, for the semi-classical 
analysis see Lipatov \cite{Lipatov}. The intercepts (binding energies) follow
to a good approximation the law 
$$\Delta_{m}= \Delta_{0}/(m+1).$$
 For the 
preferred 
$R_c=0.27\, {\rm fm}$ as chosen in 1994 in \cite{NZHERA,NZDelta} and 
supported by
the  analysis \cite{MEGGI} of lattice QCD data 
we find 
$$\Delta_{0}\equiv\Delta_{\Pom}=0.4\,.$$  
The  node of $\sigma_{1}(r)$ is located at $r=r_1\simeq 
0.056\,{\rm fm}$, for larger $m$ the rightmost node moves to a somewhat  
larger $r=r_1\sim 0.1\, {\rm fm}$. The second node of eigen-functions with
$m= 2,3$ is located at  $r_{2}\sim 3\cdot 10^{-3}~ {\rm fm}$ which corresponds
to the momentum transfer scale 
$Q^{2} ={1/r_{2}^{2}}=5\cdot 10^{3}$ GeV$^{2}$.
The third node of $\sigma_{3}(r)$ is located at $r$ beyond the reach of any
feasible DIS experiments. It has been found \cite{NZZ97} that the BFKL-Regge
 expansion (\ref{eq:3.6}) truncated at $m=2$ appears to be very successful
 in describing of the proton SFs at 
  $Q^2\lsim 200$ GeV$^2$. However, at higher $Q^2$ and
moderately
small $x\sim x_0=0.03$ the background of the CD BFKL solutions with
 smaller intercepts ($\Delta_m < 0.1$)
should be taken into account (see below).

The exchange by perturbative gluons is a dominant mechanism for small dipoles
$r\lsim R_c$. In Ref.\cite{NZHERA} interaction of large dipoles
has been modeled by the non-perturbative, soft mechanism which we
approximate here by a factorizable soft pomeron with intercept
$\alpha_{\rm soft}(0)-1=\Delta_{\rm soft}=0$ i.e. flat vs. $x$ at small $x$. The
exchange by two non-perturbative gluons has been behind the
parameterization of $\sigma_{\rm soft}(r)$ suggested in \cite{JETPVM}
and used later on in \cite{NZcharm,NZZ97,DER,PION,GamGam}.

Via equation (\ref{eq:2.4}) each hard CD BFKL eigen-cross section plus soft-pomeron 
CD cross section defines the corresponding
eigen-SF  $f_m^{c}(Q^2)$ and we arrive at the CD BFKL-Regge
expansion for the charm SF of the proton 
$(m={\rm soft},0,1,..)$ \cite{CHARM2000}
\beq
F_2^{c}(x_{Bj},Q^2)=\sum_{m} f_m^{c}(Q^2)\left(x_0\over x\right)^{\Delta_m}\,,
\label{eq:3.6}
\eeq
Now comes the crucial observation that numerically 
$r_{1} \sim  r_{S}/2$ 
and the node of hard CD BFKL eigen-cross sections is located within the peak 
of the weight function $W_{2}$. Consequently, in the calculation of open 
charm eigen-SFs $f_m^{c}(Q^2)$ one scans the eigen-cross 
section in the vicinity of the node, which leads to a strong suppression of 
subleading $f_m^{c}(Q^2)$. 
\begin{figure}[h]
\psfig{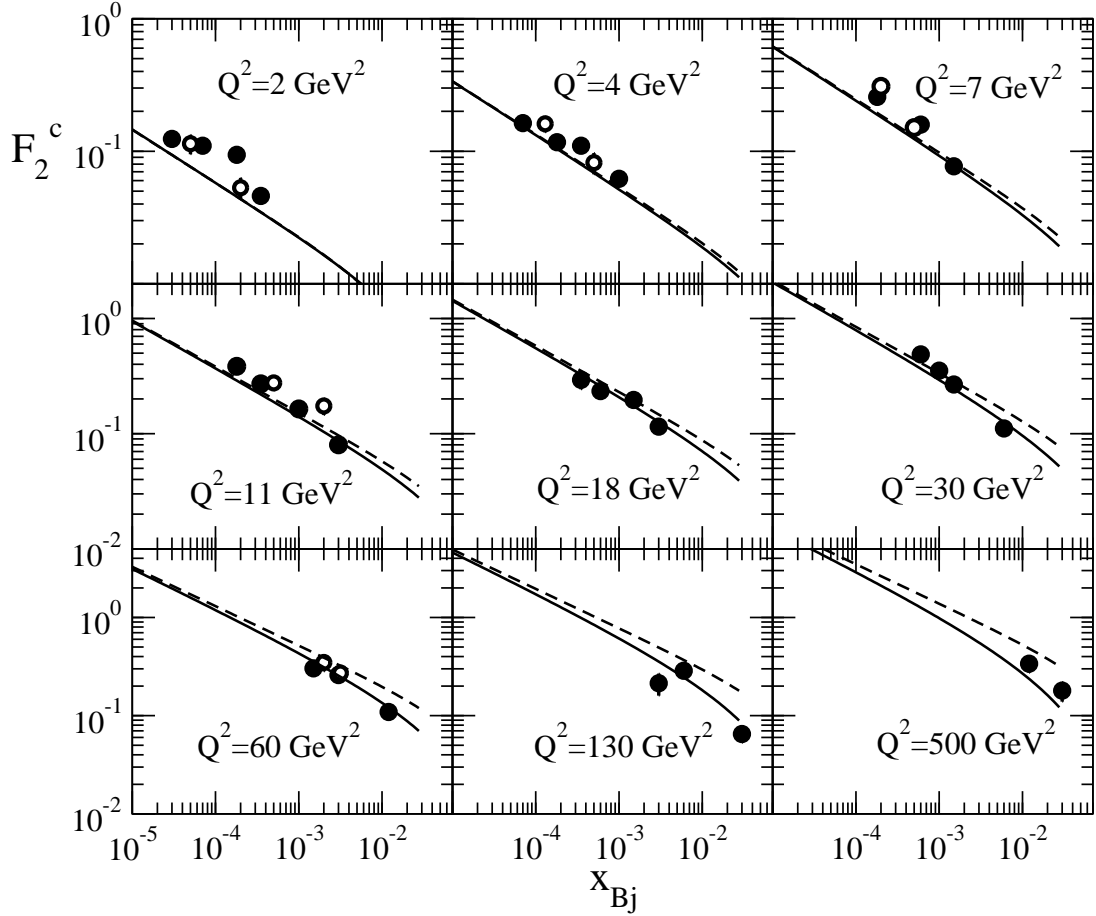}
\vspace{-0.5cm}
\caption{Prediction from CD BFKL-Regge factorization for the  charm structure
 function 
of the proton $F^{c}_2(x,Q^2)$ as a function of
the Bjorken variable $x_{Bj}$ in comparison with the experimental data from 
 H1  Collaboration \cite{H1ccbb}. The  solid curve is a result of the complete
CD BFKL-Regge expansion, the contribution of the rightmost hard BFKL pole  
 with $\Delta_{\Pom}=0.4$ is shown by dashed line}
\label{fig:fig2}
\end{figure}  

\section{Open charm 
structure functions from CD BFKL-Regge factorization}

Because a probability to find large color dipoles in the photon decreases
rapidly with the quark mass, the contribution from soft-pomeron exchange
to open charm excitation is very small down to $Q^{2}=0$.  
As we discussed elsewhere \cite{PION,CHARM2000}, for still higher solutions, 
$m\geq 3$, 
all intercepts are very small anyway, $\Delta_m\ll \Delta_{0}$,
For
this reason, for the purposes of practical phenomenology, we  truncate 
expansion (\ref{eq:3.6}) at $m=3$ lumping in the term $m=3$ contributions 
of still higher singularities with $m\geq 3$. The term $m=3$
is  endowed
 with the effective intercept $\Delta_3=0.06$
 and is presented in \cite{CHARM2000}
 in its analytical form.

We comment first on the results on $F_2^c$. The solid curve in Fig. \ref{fig:fig2}
is a result of the complete CD BFKL-Regge expansion. The dashed 
curve is the pure rightmost hard BFKL pomeron contribution (LHA), 
There is a strong cancellation between soft and  subleading 
contributions with $m=1$ and $m=3$. Consequently, for this dynamical reason 
in this region of $Q^{2} \lsim 10$ GeV$^{2}$ we have an effective one-pole 
picture and LHA gives reasonable description of  $F_2^c$.

  In agreement with the nodal structure of subleading
eigen-SFs discussed in \cite{PION,CHARM2000}, LHA over-predicts slightly 
 $F_2^c$ at $Q^{2} \gsim 
30$ GeV$^{2}$, where the negative valued subleading hard BFKL exchanges
overtake the soft-pomeron exchange  and the background from 
subleading hard BFKL exchanges becomes substantial at $Q^{2} \gsim 30$ GeV$^2$
and even the dominant component of $F_2^c$ at $Q^{2}\gsim 200$
GeV$^{2}$ and $x\gsim 10^{-2}$. In this region of $Q^{2}$ the soft-pomeron
exchange is numerically very  small.
We predicted in \cite{CHARM2000} that open charm SF
is dominated entirely by the contribution from the rightmost hard BFKL pole
at $Q^{2} \lsim 20$ GeV$^{2}$, which is due to strong cancellations 
between the soft-pomeron and subleading hard BFKL exchanges.
The soft-subleading cancellations become less accurate at smaller $x$, but
at smaller $x$ the both soft and subleading hard BFKL exchange become 
rapidly Regge suppressed $\propto x^{\Delta_{\Pom}}, 
~x^{\Delta_{\Pom}/2}$, respectively. 
\begin{figure}[h]
\psfig{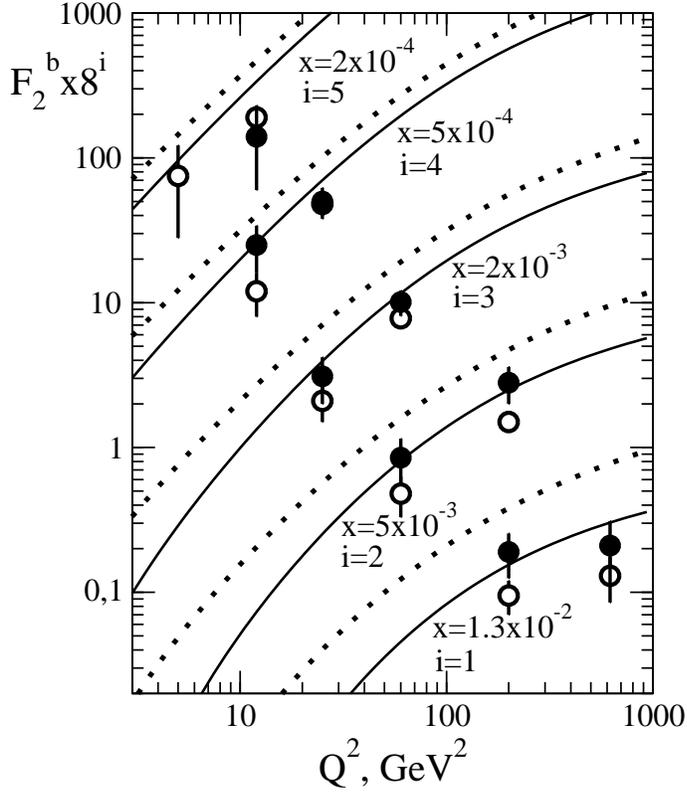}
\vspace{-0.5cm}
\caption{Comparison of predictions from CD BFKL-Regge factorization 
for the beauty 
structure function $F^{b}_2(x,Q^2)$ with data \cite{H1ccbb} (full circles) and 
\cite{H12008} (open circles).The  solid curve is a result of the complete
CD BFKL-Regge expansion, the contribution of the rightmost hard BFKL pole  
with $\Delta_{\Pom}=0.4$ is shown by dotted curves.} 
\label{fig:fig3}
\end{figure} 
In Fig.~\ref{fig:fig2} we compare our CD BFKL-Regge predictions to the 
recent experimental data from the H1 Collaboration \cite{H1ccbb} and 
find very good agreement between theory and experiment which
lends support to our 1994 evaluation $\Delta_{\Pom}=0.4$ of the 
intercept of the rightmost hard BFKL pole in the color dipole approach 
with running strong coupling. The negative 
valued contribution from subleading hard BFKL exchange is important for
bringing the theory to agreement with the experiment at large $Q^{2}$.
 For an alternative interpretation of heavy flavor production
see \cite{TOLYA, White, Baranov} and references therein.

\section{$F_2^b$ and hierarchy of pre-asymptotic pomeron intercepts }
The characteristic feature of the QCD pomeron
 dynamics at distances $\sim m_b^{-1}$ is large negative valued 
 contribution to
$F_2^b$,  coming from subleading BFKL singularities, 
see Fig. \ref{fig:fig1} and Ref.\cite{BEAUTY}. 
 Consequences of this observation
 for the exponent of the
 energy dependence of the structure function 
\beq
F_2^b\propto \left(x_0\over x\right)^{\Delta_{\rm eff}}
\label{eq:SBBEXP}
\eeq
 are quite interesting.
In terms of the ratio
$r_m=\sigma_m/\sigma_0$ (see Fig. {\ref{fig:fig1})
 the exponent
$\Delta_{\rm eff}$ reads (m=1,2,3,soft)\cite{BEAUTY}
\beq
\Delta_{\rm eff}=\Delta_{0}\left[1-\sum_{m=1}r_m(1-\Delta_m/\Delta_0)
({x_0/ x})^{\Delta_m-\Delta_0}\right]
\label{eq:DELTAEFF}
\eeq
 Coefficients $r_m$
 in eq.(\ref{eq:DELTAEFF}) depend on $r$. They are negative on the left from
the rightmost node (Fig. \ref{fig:fig1}) and positive on the right.
Because for  $r\sim m_b^{-1}$ 
all $r_m$
are negative, except $r_{\rm soft}(0)>0$ \cite{BEAUTY}, at HERA energies
 the effective
 intercept $\Delta_{\rm eff}\equiv \Delta_{\rm beauty}$ 
overshoots the asymptotic value
 $$\Delta_{\Pom}\equiv \Delta_0=0.4\,.$$
 At still higher collision energies  both  the soft and subleading hard BFKL
  exchanges become rapidly Regge suppressed. This results in
 decreasing $\Delta_{\rm eff}$ down to
 $\Delta_{\Pom}$ \cite{BEAUTY}.

For comparison,
 in photoproduction of open charm which scans, as we discussed above,
 the color dipole cross section at distances
$\sim 1/m_c$, in the vicinity of the rightmost node,
 there is a strong cancellation between soft and  subleading
contributions to $F_2^{c}$ \cite{NZcharm,CHARM2000}.
 Consequently, for this dynamical
 reason in open  charm photoproduction we have an effective one-pole
picture and the effective pomeron intercept
 $$\Delta_{\rm eff}\equiv\Delta_{\rm charm} \simeq \Delta_{\Pom}$$.

In photoproduction of light flavors the CD cross section $\sigma_m(r)$ 
is close to the
 saturation regime $\sigma_m(r)\propto const$
and all subleading and soft terms of the CD BFKL-Regge expansion
 are positive valued
and numerically important (see \cite{PION} for more details).
This is the dynamical reason for smallness of a pre-asymptotic
 pomeron intercept in  photoproduction of light flavors.
Hence,   the  hierarchy of  pre-asymptotic 
intercepts 
\beq
\Delta_{\rm beauty}>\Delta_{\rm charm}>\Delta_{\rm light}
\label{eq:BCL}
\eeq
which
brings to light the internal dynamics of leading-subleading cancellations at
different hardness scales. 

In Fig. \ref{fig:fig3}  we presented our predictions for the beauty 
structure function.
 The solid curve corresponds to the
complete expansion (\ref{eq:3.6}) while 
 the dotted
curve is the leading hard pole approximation, $F_2^{b}(x,Q^2)\simeq
 f_0^{b}(Q^2)\left(x_0/ x\right)^{\Delta_0}$.
 In agreement with the nodal structure of subleading
eigen-SFs the latter  over-predicts
 $F_2^b$ significantly because the negative valued contribution from
 subleading hard BFKL
 exchanges
overtakes the soft-pomeron exchange and the background from
subleading hard BFKL exchanges is substantial for all $Q^{2}$ \cite{BEAUTY}.
Our structure functions obtained  in 1999
 agree well  with the  determination of  $F_2^b$  
by the H1 published in 2006 \cite{H1ccbb} (full circles) but  contradict to very 
recent (2008, preliminary) 
H1 results on $F_2^b$ \cite{H12008} (open circles).
\begin{figure}[h]
\psfig{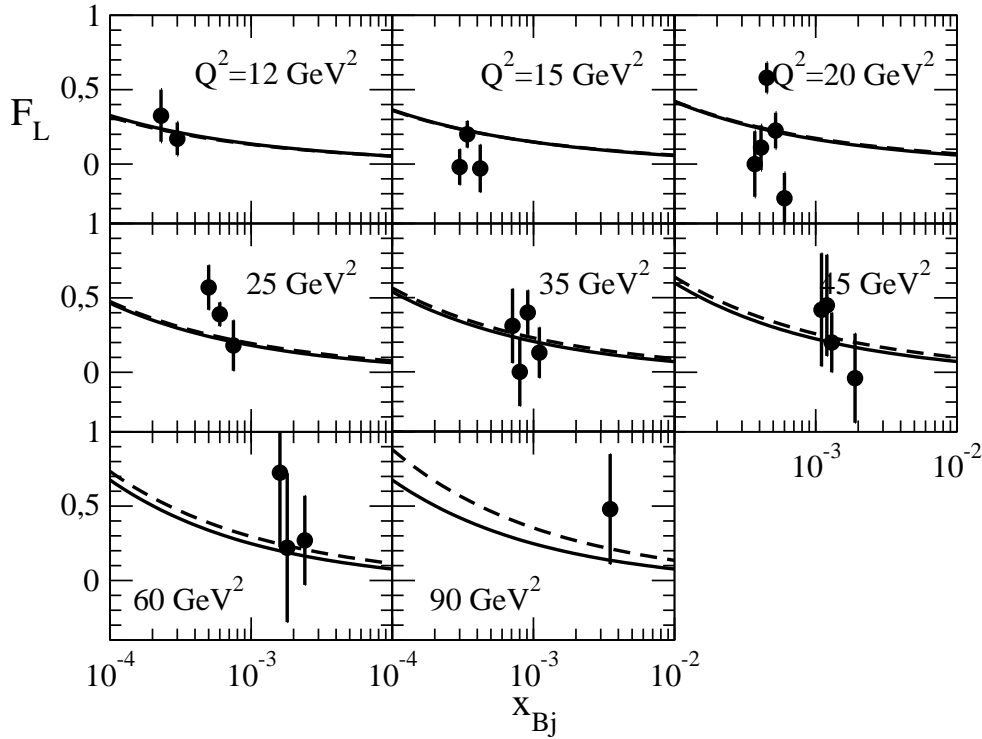}
\caption{Prediction from CD BFKL-Regge factorization for the longitudinal 
structure function 
of the proton $F_L(x,Q^2)$ as a function of
the Bjorken variable $x_{Bj}$. The  solid curve is a result of the complete
CD BFKL-Regge expansion, the contribution of the rightmost hard BFKL pole  
with $\Delta_{\Pom}=0.4$ is shown by dashed line. Data points are from \cite{H1FL2008}} 
\label{fig:fig4}
\end{figure}  
\section{ Elastic $\Upsilon(1S)$ meson
photoproduction.}

The cross section of elastic $\Upsilon(1S)$ meson
photoproduction has been measured at HERA \cite{ELBEAUTY}. Quarks in
 $\Upsilon$ meson are nonrelativistic and $|\gamma\rangle\propto
 m_bK_0(m_br)$. The forward $\gamma\to \Upsilon$ transition matrix element
 $\langle\Upsilon|\sigma_n(r)|\gamma\rangle$ is controlled by the
 product $\sigma_0(r)K_0(m_br)$ \cite{SCAN}
and the amplitude of  elastic
 of $\Upsilon(1S)$ photoproduction
  is dominated by the contribution from the dipole
sizes $r\sim r_{\Upsilon}=A/m_{\Upsilon}$
with $A= 5$. The crucial observation   is that at distances 
$r\sim r_{\Upsilon}$
  cancellation between soft and subleading contributions to the elastic
photoproduction cross section  results in the exponent
  $\Delta$ in 
\beq
{d\sigma(\gamma p\to\Upsilon p )\over dt}|_{t=0}\propto W^{4\Delta}
\label{eq:ELBB}
\eeq
 which is very close to $\Delta_{\Pom}$,  $\Delta=0.38$ \cite{BEAUTY,JETPVM}. This
observation  appears  to be in
agreement with the cross section rise observed by ZEUS\&H1 
\cite{ELBEAUTY}.

\section{$\Delta_{\Pom}$ from
measurements of $F_L(x,Q^2)$}

It has been demonstrated in \cite{NZDelta} that the longitudinal structure
function $F_L(x,Q^2)$ emerges as local probe of the dipole
cross section at $r^2\simeq 11./Q^2$.
The subleading CD BFKL cross sections have their rightmost node
at $r_1\sim 0.05-0.1$ fm. Therefore, one can zoom at the leading CD BFKL  pole
contribution and measure the pomeron intercept
$\Delta_{\Pom}$ from the $x$-dependence of $F_L(x,Q^2)$
at $Q^2\sim 10-30$ GeV$^2$. 
The discussed above cancellation of the  soft-subleading contributions
 is nearly exact at $Q^2\sim 10-30$ GeV$^2$. This results in the leading 
hard pole dominance in this region,
see Fig. \ref{fig:fig4}) where comparison with the very recent 
H1 data \cite{H1FL2008} is presented.

\section{Electromagnetic corrections to the Okun-Pomeranchuk theorem}

Compare  the total cross section of  charged and neutral components of
isotriplets of mesons like the $\rho$-meons or pions  on an electrically
neutral target like a neutron. Arguably, for such a target the electromagnetic
breaking of the Okun-Pomeranchuk theorem \cite{OP} will be dominated by the
electromagnetic lifting of the  degeneracy of sizes of charged and neutral
$\rho$'s. The strength of the Coulomb interactioin in the charge and neutral
mesons is proportional to $e_u e_d$ and $-(e_u^2 +e_d^2)/2$,  respectively, the
net difference is $\propto (e_u+e_d)^2$. Consequently, the difference of the
radii mean squared can be estimated as $\sim  \alpha_{em}(e_u+e_d)^2\langle r^2
\rangle$, what would entail
\beq
{{\sigma_{\pm} - \sigma_o } \over {\sigma_{\pm} + \sigma_o }} \sim 
\alpha_{em}(e_u+e_d)^2
\label{eq:OKUN1}
\eeq

\section{Conclusions}

Color dipole approach to the BFKL dynamics predicts uniquely decoupling 
of subleading hard BFKL exchanges from open charm SF of
the proton at $Q^2\lsim 20\,{\rm GeV^2}$, from $F_L$ at $Q^2\simeq 20\,{\rm GeV^2}$ 
and from $\partial F_2/\partial \log Q^2$ at 
 $Q^2\simeq 4\,{\rm GeV^2}$. This decoupling is due to
dynamical cancellations between contributions of different subleading
hard BFKL poles and leaves us with an effective soft+rightmost hard BFKL 
two-pole approximation with intercept of the 
soft pomeron $\Delta_{{\rm soft}}=0$. 
We predict strong cancellation between the soft-pomeron 
and subleading hard BFKL contribution to $F_2^c$ in the experimentally
 interesting region
of $ Q^{2} \lsim 20$ GeV$^{2}$, in which $F_2^c$
is dominated entirely by the contribution from the rightmost hard BFKL pole.
This makes open charm in DIS at $Q^2\lsim 20$ GeV$^{2}$ a unique 
handle on the intercept of the rightmost hard BFKL exchange.

High-energy open beauty
photoproduction probes the vacuum exchange at  distances $\sim 1/m_b$
and detects significant  corrections to the 
BFKL asymptotics coming from the  subleading  vacuum poles.
 We show that the
interplay of leading and subleading vacuum exchanges gives rise to the
cross section $\sigma^{b\bar b}(W)$ growing much faster than it is
prescribed by the exchange of the leading pomeron trajectory with
intercept $\alpha_{\Pom}(0)-1=\Delta_{\Pom}=0.4$. Our calculations
within the  color dipole BFKL model are in
agreement with the recent determination of $\sigma^{b\bar b}(W)$ by the H1
collaboration.  The comparative analysis
 of diffractive photoproduction of  beauty, charm and light quarks exhibits
the hierarchy of pre-asymptotic pomeron intercepts  which follows the
hierarchy of corresponding hardness scales.
  We comment  on the phenomenon of decoupling of soft and
subleading BFKL singularities at the scale of elastic $\Upsilon(1S)$
 -photoproduction which results in precocious color dipole
 BFKL asymptotics of the
process $\gamma p \to \Upsilon p$.

Similar hard BFKL pole dominance holds for $F_L(x,Q^2)$ .  
At still higher values of $Q^{2}$ the soft-pomeron exchange is predicted to
die out and negative valued background contribution from subleading hard BFKL
exchange with effective  intercept $\Delta_{3}\approx 0.06$ 
becomes substantial at not too small $x\sim x_0$. The agreement with the 
presently available
experimental data on open charm/beauty  in DIS confirm
 the CD BFKL prediction of the intercept $\Delta_{\Pom}=0.4$ 
for the rightmost hard BFKL-Regge pole. The experimental 
confirmation of our predictions for hierarchy 
of soft-hard exchanges as function of $Q^{2}$ 
is  a strong argument in favor of the CD BFKL
approach. \\

{\bf Acknowledgments: } This contributions is a humble tribute to  Lev 
Borisovich Okun on 
the occasion of his 80th birthday. The authors benefited tremendously 
from the interaction 
with LBO as a diploma and PhD advisor in the case of NNN and as a collegaue 
at ITEP in the  case of VRZ.

The work was supported in part  by
 the RFBR grant  07-02-00021.



\end{document}